\documentclass[journal=jctcce,manuscript=article,layout=twocolumn]{achemso}
\usepackage{bm}
\usepackage{physics}
\usepackage[utf8]{inputenc}
\usepackage[T1]{fontenc}
\usepackage{newtx} 
\usepackage{dsfont} 
\usepackage{braket}

\makeatletter
\let\l@addto@macro\relax
\makeatother
\usepackage[fontsize=11pt]{scrextend}

\author{Ken Miyazaki}
\author{Alex Krotz}
\author{Roel Tempelaar}
\email{roel.tempelaar@northwestern.edu}
\affiliation{Department of Chemistry, Northwestern University, 2145 Sheridan Road, Evanston, Illinois 60208, USA}

\title{Unitary Basis Transformations in Mixed Quantum--Classical Dynamics}

\begin{document}

\begin{tocentry}
\centering
\includegraphics{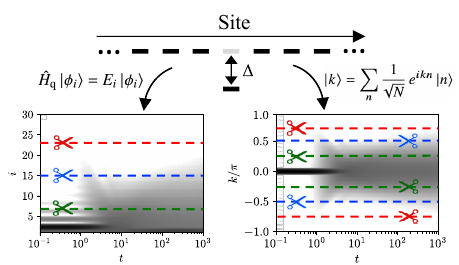}
\end{tocentry}

\maketitle

\begin{abstract}
A common approach to minimizing the cost of quantum computations is by transforming a quantum system into a basis that can be optimally truncated. Here, we derive classical equations of motion subjected to similar unitary transformations, and propose their integration into mixed quantum--classical dynamics, enabling this class of methods to be applied within arbitrary bases for both the quantum and classical coordinates. To this end, canonical positions and momenta are combined into a set of complex-valued classical coordinates amenable to unitary transformations. We demonstrate the potential of the resulting approach by means of surface hopping calculations of an electronic carrier scattering onto a single impurity in the presence of phonons. Appropriate basis transformations, capturing both the localization of the impurity and the delocalization of higher-energy excitations, are shown to faithfully capture the dynamics within a fraction of the classical and quantum basis sets.
\end{abstract}

\section{Introduction}\label{sec:intro}

The application of unitary basis transformations is common practice in quantum-mechanical computations, derivations, and analyses. Computations typically require bases to be truncated, and depending on the quantum system at hand, the effectiveness of such truncations varies with the transformed representation \cite{taghizadeh2017, taylor2020}. In addition, transformations enable one to construct perturbative expansions with optimal convergence properties, while also allowing quantum-mechanical equations to be cast in their most intuitive form.

This is exemplified by Bloch's theorem \cite{bloch1929}, which finds widespread application in the modeling of materials. Bloch's theorem invokes a complex Fourier transform of physical basis states over a crystal lattice in order to yield a representation in reciprocal space \cite{kittel1966}. This representation naturally captures the conservation of lattice momentum, and describes phenomena in terms of quasiparticle bands that can be included or excluded in order to modulate the computational cost \cite{taghizadeh2017}. Another example is provided by Redfield theory \cite{redfield1965}, which is formulated within the eigenbasis of a quantum system, thereby allowing the nonadiabatic coupling between (adiabatic) eigenstates to be captured perturbatively. Here, the computational cost can be modulated by including or excluding eigenstates \cite{tempelaar2018}. More generally, for a quantum system expressed in a physical basis (also referred to as local basis or ``site'' basis), an arbitrary unitary transformation takes the form
\begin{align}
    \ket{\xi} = \sum_n u_{\xi n} \ket{n},
    \label{eq:n_to_xi}
\end{align}
with $u \in \mathbb{C}$ and $u^\dagger u = \mathds{1}$, and where $n$ and $\xi$ label the physical and transformed basis states, respectively.

In classical mechanics, basis transformations take the form of canonical transformations. For a given set of canonical position and momentum coordinates, denoted $\{q_n\}$ and $\{p_n\}$, respectively, the transformed canonical coordinates take the form $q_\xi = q_\xi(\{q_n, p_n\})$ and $p_\xi = p_\xi(\{q_n, p_n\})$. These transformed coordinates serve the role of ``position'' and ``momentum'' in the transformed Hamilton equations of motion, respectively, although they do not necessarily correspond to a physical position and momentum. All coordinates are real valued, as is required for classical trajectories. Canonical transformations offer the same benefits to classical mechanics as unitary transformations do to quantum mechanics, yet the interconnection between canonical and unitary transformations is somewhat opaque. Solidifying this interconnection is of particular relevance to mixed quantum--classical (MQC) dynamics, where electronic excitations are described quantum-mechanically and nuclear vibrations (phonons) classically.

MQC dynamics (sometimes referred to as nonadiabatic molecular dynamics) finds widespread application to the transient modeling of excited-state phenomena in molecular systems \cite{nelson2014, subotnik2016, wang2016, crespo-otero2018, nelson2020}. Virtually all implementations of MQC dynamics adopt a physical basis for the nuclear coordinates. This practice is perhaps motivated by the notion that molecular excited-state phenomena are commonly localized, with site-to-site energy transfer being mediated by local vibrations. Such is indeed captured most efficiently by a physical basis, as localization allows this basis to be truncated to only the spatial regions of interest.

In a recent work \cite{alex2021}, henceforth referred to as Paper I, we derived a formulation of MQC dynamics fully within reciprocal space by subjecting both quantum and classical coordinates to a complex Fourier transform. This work was motivated by a surge in applications of MQC dynamics to crystalline materials \cite{nie2014, nie2015, long2016, chu2016, zheng2018, shi2020, smith2019, zhang2021, xie2022, lively2024}, for which excited-state phenomena are typically delocalized and driven by an exchange of lattice momentum between electronic carriers and Bloch-like phonons \cite{mahan2013}. Arriving at the reciprocal-space MQC formalism required us to first combine classical positions and momenta within a single complex-valued coordinate, $z_n$, amenable to a complex Fourier transform. The transformed equations of motion were shown to yield dynamics formally-equivalent to solutions in the physical basis. At the same time, the reciprocal-space representation was shown to allow for basis truncations of band-like dynamics not possible within a physical basis representation, allowing the computational cost to be optimally reduced. In Paper I, reciprocal-space MQC was introduced for mean-field dynamics (sometimes referred to as Ehrenfest dynamics). In a follow-up work, henceforth referred to as Paper II \cite{alex2022}, we have extended reciprocal-space MQC for the popular fewest-switches surface hopping (FSSH) method \cite{tully1990}. This formulation has since found application in the modeling of the Floquet nonadiabatic dynamics of laser-dressed solid-state materials \cite{chen2024}.

While reciprocal-space MQC is particularly effective in describing band-like phenomena, its effectiveness deteriorates once the periodicity of the crystal lattice becomes disrupted, and lattice momentum is no longer a good quantum number. Such disruptions may take the form of defects, such as impurities, vacancies, and dislocations. In such cases, the physical basis may not provide an effective representation either, as excited states may retain substantial delocalization lengths. Instead, the optimal representation will be provided by some other basis. More generally, for \emph{any} given system, out of all possible bases one should be able to find a choice that most effectively captures the dynamics. In that regard, localized molecular excitations and band-like excitations in materials span the two Fourier-related extremes, captured optimally by the physical and reciprocal bases, respectively.

In this Article, we introduce a formulation of MQC dynamics within \textit{arbitrary} basis representations for both quantum and classical coordinates, opening the opportunity to transiently model excited-state phenomena within optimized bases for \textit{any} given system. As in Paper I \cite{alex2021}, we combine classical positions and momenta within a complex-valued coordinate, $z_{n}$, which is subjected to an arbitrary basis transformation, similarly to the quantum basis states in Eq.~\ref{eq:n_to_xi}. We then derive the transformed equations of motion, which are integrated in MQC dynamics. In order to demonstrate the utility of this framework, we consider a model invoking an electronic carrier scattering onto a single impurity in an otherwise pristine lattice, under Holstein-type coupling to phonons. For this model, fewest-switches surface hopping (FSSH) calculations are presented, showing the excited state dynamics to be invariant to significant truncations of the appropriately-transformed electronic and nuclear bases.

This Paper is organized as follows. In Sec.~\ref{sec:theory} we introduce the transformed classical equations of motion, their integration in MQC dynamics, and in FSSH in particular. In Sec.~\ref{sec:recip}, we use the resulting formalism to re-derive the equations of motion of reciprocal-space MQC and FSSH from Papers I \cite{alex2021} and II \cite{alex2022}. In Sec.~\ref{sec:impurity} we introduce the single impurity model, and present results from transformed FSSH under basis truncations. In Sec.~\ref{sec:conclusions}, we present our conclusions and offer an outlook for future directions.

\section{Theory\label{sec:theory}}

Since unitary basis transformations of quantum systems are trivial and well-established, we instead begin by considering such transformations for classical systems, before presenting the integration of the resultant classical equations of motion within MQC dynamics.

\subsection{Transformed classical dynamics}\label{sec:classical}

\begin{figure}[t!]
    \centering
    \includegraphics{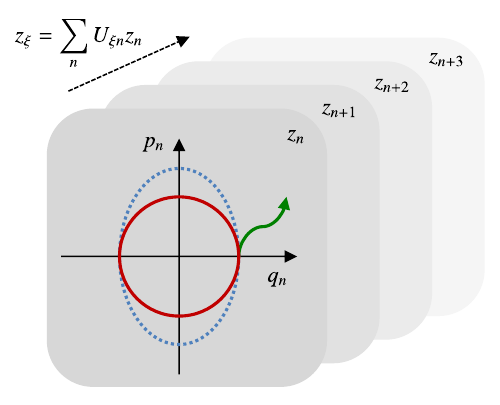}
    \caption{Schematic depiction of transformed classical dynamics. The canonical position and momentum, $q_{n}$ and $p_{n}$, are combined into a complex-valued coordinate, $z_{n}$, following Eq.~\ref{eq:gen_z}. The relative scaling of $q_{n}$ and $p_{n}$ along the real and imaginary axes is modulated by $h_n$ (see text), such that harmonic motion is mapped onto a perfect circle when $h_n = \omega$ (red solid curve) while being mapped onto an ellipse when $h_n \neq \omega$ (blue dash). Arbitrary classical trajectories can be represented within this construction (green solid curve). The resulting set of complex-valued coordinates can then be subjected to arbitrary unitary basis transformations, producing transformed coordinates, $z_{\xi}$.}
    \label{fig:scheme}
\end{figure}

The general idea behind transformed classical dynamics is schematically depicted in Fig.~\ref{fig:scheme}. Here, a set of canonical position and momentum coordinates is introduced, which are described by a generic Hamiltonian function of the form
\begin{align}
    H(\{q_{n},p_{n}\}) = \sum_{n}\frac{p_{n}^2}{2m_{n}} + V(\{q_{n}\}).
\end{align}
Here, the first term represents kinetic energy, where $m_{n}$ denotes the mass of coordinate $n$, and the second term represents potential energy. The time evolution of the canonical position and momentum coordinates is governed by the Hamilton equations of motion,
\begin{align}
    \label{eq:HamiltonsEOM}
    \dot{q}_{n} = \frac{\partial H}{\partial p_{n}} = \frac{p_{n}}{m_{n}}, \quad \dot{p}_{n} = -\frac{\partial H}{\partial q_{n}} = -\frac{\partial V}{\partial q_{n}}.
\end{align}

As in Paper I \cite{alex2021} and as illustrated in Fig.~\ref{fig:scheme}, we combine the classical position and momentum within a single complex-valued coordinate for each $n$,
\begin{align}
    \label{eq:gen_z}
    z_{n} = \sqrt{\frac{m_{n}h_{n}}{2}} \left(q_{n} + i\frac{p_{n}}{m_{n}h_{n}}\right).
\end{align}
Here, we introduced a variable $h_n$ whose magnitude is arbitrary, and which modulates the scaling of the position and momentum along the real and imaginary axes in the complex plane, respectively. Expressed in terms of the complex-valued coordinates, the generic Hamiltonian function is given by
\begin{align}
    \label{eq:Hz}
    H(\{z_{n}\}) = -\sum_{n} \frac{h_{n}}{4} \, \left( z_{n}^{2} -2\,z_{n}z_{n}^{*} + z_{n}^{*2} \right) + V(\{z_{n}\}).
\end{align}The corresponding Hamilton equations take the form
\begin{align}
    \dot{z}_{n} = -i \frac{\partial H}{\partial z_{n}^{*}}.
\end{align}

In Paper I \cite{alex2021}, the classical coordinates were assumed to represent identical harmonic modes, meaning that $m_n = m$ and
\begin{align}
    V(\{q_{n}\}) = \frac{1}{2} m \omega^2 \sum_n q_n^2.
    \label{eq:V_harmonic}
\end{align}
The harmonic frequency $\omega$ then took the place of $h_n$ in Eq.~\ref{eq:gen_z}. This choice of $h_n$ maps harmonic motion onto a perfect circle in the complex plane, as shown in Fig.~\ref{fig:scheme}. As a result, the Hamiltonian function simplifies to $H(\{z_n\}) = \omega \sum_{n} z_{n} z_{n}^{*}$ and the associated time evolution follows as $\dot{z}_{n} = -i \omega \, z_{n}$. In that case, $z_n$ may be interpreted as the eigenvalue associated with the coherent state of the harmonic oscillator at coordinate $n$ \cite{kim2022}. When $h_n \neq \omega$, harmonic motion is mapped onto an ellipse instead. In the present analysis, no assumption is made about the modes, and by leaving $h_n$ unspecified we derive equations of motion applicable under generic potentials, including anharmonicities.

With the positions and momenta combined into a set of complex coordinates, arbitrary unitary basis transformations can be applied following
\begin{align}
    \label{eq:z_xi}
    z_{\xi} = \sum_{n} U_{\xi n} z_{n},
\end{align}
as depicted in Fig.~\ref{fig:scheme}. Here, $U \in \mathbb{C}$ and $U^\dagger U = \mathds{1}$. Eq.~\ref{eq:z_xi} is the classical equivalent of Eq.~\ref{eq:n_to_xi}. As can be easily verified, the Hamilton equations expressed in terms of the transformed complex-valued coordinates follow as
\begin{align}
    \label{eq:z_xi_dot}
    \dot{z}_{\xi} = -i \, \pdv{H}{z_{\xi}^{*}}.
\end{align} The transformed Hamiltonian function, on the other hand, takes the slightly more complicated form
\begin{align}
    H(\{z_{\xi}\}) =& -\frac{1}{4} \, \sum_{\xi,\xi'} \big( \tilde{h}^{*}_{\xi \xi'} z_{\xi} z_{\xi'} -2 h_{\xi \xi'} z_{\xi} z_{\xi'}^{*} \nonumber \\
    & + \tilde{h}_{\xi \xi'} z_{\xi}^{*} z_{\xi'}^{*} \big) + V(\{z_{\xi}\}),
    \label{eq:H_xi}
\end{align}
where $h_{\xi \xi'} \equiv \sum_{n} U^{*}_{\xi n} h_{n} U_{\xi' n}$ and $\tilde{h}_{\xi \xi'} \equiv \sum_{n} U_{\xi n} h_{n} U_{\xi' n}$.

Eq.~\ref{eq:Hz} provides a generic treatment of any classical system by appropriate construction of the potential energy in terms of the complex-valued coordinates, $V(\{z_n\})$. Eq.~\ref{eq:H_xi} is formally equivalent to Eq.~\ref{eq:Hz}, and the associated equations of motion given by Eq.~\ref{eq:z_xi_dot} allow the system's time evolution to be determined within an arbitrary basis. Notably, the transformed complex-valued coordinates $z_{\xi}$ can be decomposed into real-valued canonical coordinates $q_{\xi}$ and $p_{\xi}$, for which the Hamilton equations of motion can be derived straightforwardly. This establishes the connection between the unitary transformation applied in Eq.~\ref{eq:z_xi} and canonical transformations of the form $q_\xi = q_\xi(\{q_n, p_n\})$ and $p_\xi = p_\xi(\{q_n, p_n\})$. In the following, however, we will minimize such analyses, and instead resort to the complex-valued coordinates, which provide a simpler and general framework for describing classical dynamics. This has the added benefit of putting classical basis transformations on the same footing as quantum-mechanical basis transformations.

\subsection{Mixed quantum--classical dynamics}\label{sec:MQC}

We now proceed to integrate the transformed classical coordinates within MQC dynamics. This is not intended as a rigorous introduction into MQC dynamics, for which we refer the interested reader to excellent sources in the literature \cite{nelson2014, subotnik2016, wang2016, crespo-otero2018}.

MQC dynamics relies on the subdivision of a system of interest into a quantum subsystem and a classical subsystem, where the quantum subsystem is commonly taken to represent the electronic states while the classical subsystem is reserved for nuclear coordinates. Resorting to canonical coordinates within the physical basis, the total Hamiltonian takes the form
\begin{align}
    \label{eq:H_mqc_physical}
    \hat{H}_{\mathrm{tot}}(\{q_{n}, p_{n}\}) = \hat{H}_{\mathrm{q}} + \hat{H}_{\mathrm{q-c}}(\{q_{n}\}) + H_{\mathrm{c}}(\{q_{n}, p_{n}\}).
\end{align}
Here, $\hat{H}_{\mathrm{q}}$ is the Hamiltonian operator of the quantum subsystem, $H_{\mathrm{c}}(\{q_{n}, p_{n}\})$ is the Hamiltonian function of the classical subsystem, and $\hat{H}_{\mathrm{q-c}}(\{q_{n}\})$ is the operator governing interactions between the quantum and classical subsystems, which involves a parametric dependence on the physical position coordinates. Adopting transformed coordinates, this yields
\begin{align}
    \label{eq:H_mqc}
    \hat{H}_{\mathrm{tot}}(\{z_{\xi}\}) = \hat{H}_{\mathrm{q}} + \hat{H}_{\mathrm{q-c}}(\{z_{\xi}\}) + H_{\mathrm{c}}(\{z_{\xi}\}).
\end{align}

The quantum--classical interaction term contributes to the evolution of the quantum subsystem, which is governed by the time-dependent Schr\"odinger equation,
\begin{align}
    i\hbar\,\ket{\dot{\Psi}} = \left(\hat{H}_{\mathrm{q}} + \hat{H}_{\mathrm{q-c}}(\{z_{\xi}\})\right) \ket{\Psi},
\end{align}
where $\Psi$ is the quantum wavefunction. It also affects the evolution of the classical subsystem, since the \emph{total} Hamiltonian function to be used in the Hamilton equation for the classical coordinates (Eq.~\ref{eq:H_xi}) receives potential energy contributions from the quantum--classical interaction as well as those intrinsic to the classical subsystem. That is,
\begin{align}
    V(\{z_\xi\}) = V_\mathrm{c}(\{z_\xi\}) + V_\mathrm{q-c}(\{z_\xi\}),
\end{align}
where the quantum--classical contribution is given by some expectation value of the quantum--classical Hamiltonian, $V_{\mathrm{q-c}}(\{z_\xi\}) = \braket{\hat{H}_{\mathrm{q-c}}(\{z_{\xi}\})}$.

Various MQC dynamical methods differ in the way this expectation value is determined. In case of Ehrenfest dynamics, where the quantum--classical interaction is treated as a mean-field problem, the expectation value is taken with respect to the quantum wavefunction, yielding $\braket{\hat{H}_{\mathrm{q-c}}(\{z_{\xi}\})} = \braket{\Psi | \hat{H}_{\mathrm{q-c}}(\{z_{\xi}\}) | \Psi}$. For FSSH, on the other hand, the expectation value is taken based on an instantaneous eigenstate $\alpha$ of the total Hamiltonian operator, satisfying
\begin{align}
    \left(\hat{H}_\mathrm{q} + \hat{H}_\mathrm{q-c}\right) \ket{\alpha} = \epsilon_\alpha \ket{\alpha}.
    \label{eq:SE_full}
\end{align}
That is, a single ``active surface'', denoted $a$, is chosen and the expectation value is determined as $\braket{\hat{H}_{\mathrm{q-c}}(\{z_{\xi}\})} = \braket{a | \hat{H}_{\mathrm{q-c}}(\{z_{\xi}\}) | a}$.

A key ingredient of FSSH is a stochastic switching of the active surface between instantaneous eigenstates. The switching from state $\alpha$ to state $\beta$ is governed by the probability \cite{tully1990, sharon1994, alex2022}
\begin{align}
    \label{eq:probability}
    P_{a:\alpha\rightarrow\beta} = 2\Re \left( \Braket{\alpha|\frac{\partial \beta}{\partial t}} \frac{A_\beta}{A_\alpha} \right)\Delta t.
\end{align}
Here, $\Delta t$ is the time increment for which the switching probability is evaluated, and $A_\alpha$ is the coefficient of $\Psi$ expanded in the instantaneous eigenbasis, i.e.,
\begin{align}
    \ket{\Psi} = \sum_\alpha A_\alpha \ket{\alpha}.
\end{align}
We note that the switching probability commonly features a product of momentum and the nonadiabatic coupling vectors within the physical basis \cite{tully1990}. By means of the chain rule \cite{sharon1994, alex2022}, we have replaced this product by an inner product of $\alpha$ and $\partial \beta/\partial t$ in Eq.~\ref{eq:probability}, yielding a basis-independent form of the switching probability.

Upon a switch, the conservation of total (quantum plus classical) energy is reinforced by a rescaling of the physical momenta of the classical subsystem in the direction of the nonadiabatic coupling vector. Accordingly,
\begin{align}
    p_{n}' = p_{n} - \gamma \Braket{\tilde{\alpha} | \frac{\partial}{\partial q_{n}} | \tilde{\beta}},
\end{align}
where $p_{n}'$ and $p_{n}$ are the new and old momentum, respectively. Here, the tilde in $\tilde{\beta}$ refers to the projection of the potentially complex-valued eigenvector $\ket{\beta}$ onto a real-valued vector, in order to ensure that the physical momentum coordinate remains real-valued, and same for $\tilde{\alpha}$. Complex values for eigenvectors may arise due to an arbitrary global phase, complex basis transformations followed by basis truncations \cite{alex2022}, and geometric phase effects \cite{miao2019, wu2021, bian2022, daggett2024}. In a recent work, we have proposed a means to perform this projection onto real-valued vectors while ensuring gauge-invariance \cite{krotz2024}.
    
Within the transformed basis, the rescaling takes the form
\begin{align}
    \label{eq:rescale}
    z_{\xi}' = z_{\xi} - i\gamma \Braket{\tilde{\alpha} | \frac{\partial}{\partial z_{\xi}^*} | \tilde{\beta}}.
\end{align}

\subsection{Reciprocal space}\label{sec:recip}

In Papers I \cite{alex2021} and II \cite{alex2022}, we derived a formulation of MQC dynamics fully within reciprocal space, and applied the resulting method to pristine one-dimensional lattice models involving a single electronic carrier interacting with harmonic nuclear vibrations. In the following, we will show that the generalized equations presented in the current work recover those from Paper I \cite{alex2021} when the unitary transformation of the classical subsystem is taken to be a complex Fourier transforms over the lattice, which underlies Bloch's theorem.
    
Accordingly, we replace $\xi$ by the wavevector $k$ (which quantifies the lattice momentum), and take
\begin{align}
    \label{eq:Fourier}
    U_{kn} = \frac{1}{\sqrt{N}}\,e^{ikn},
\end{align}
where $N$ is the total number of lattice sites. The associated classical coordinates are taken to represent identical harmonic modes, such that $m_{n} = m$ and $V_\mathrm{c}(\{q_{n}\})$ is given by Eq.~\ref{eq:V_harmonic}, with $\omega$ the mode frequency. Setting $h_{n} = \omega$, the complex-valued classical coordinates take the form
\begin{align}
    \label{eq:z_n_harmonic}
    z_{n} = \sqrt{\frac{m\omega}{2}} \left(q_{n} + i\,\frac{p_{n}}{m\omega} \right).
\end{align}
Subjecting these coordinates to the transformation given in Eq.~\ref{eq:Fourier}, we arrive at transformed coordinates $z_k$. These coordinates are associated with phonons, i.e., nuclear vibrational quasiparticles with a well-defined lattice momentum.
    
Expressed in terms of the transformed coordinates, the purely-classical potential energy contribution is given by
\begin{align}
    \label{eq:V_recip}
    V_\mathrm{c}(\{z_{k}\}) = \frac{\omega}{4} \sum_{k} \left( z_{k}z_{-k} + 2\,z_{k}z_{k}^{*} + z_{k}^{*}z_{-k}^{*} \right).
\end{align}
As a result, the Hamiltonian to be used in the Hamilton equations reduces to
\begin{align}
    \label{eq:H_recip}
    H(\{z_{k}\}) = \omega \sum_{k} z_{k}z_{k}^{*} + V_{\mathrm{q-c}}(\{z_{k}\}).
\end{align}
In accordance with Eq.~\ref{eq:z_xi_dot}, this yields
\begin{align}
    \label{eq:z_dot_recip}
    \dot{z}_{k} = -i\,\omega z_{k} - i\, \pdv{V_{\mathrm{q-c}}(\{z_{k}\})}{z_{k}^{*}}.
\end{align}

In Paper I \cite{alex2021}, canonical coordinates were reconstructed from $z_k$ following
\begin{align}
    q_{k} \equiv \sqrt{\frac{2}{m\omega}} \Re{z_{k}}, \quad
    p_{k} \equiv \sqrt{2m\omega} \Im{z_{k}},
\end{align}
which obeys
\begin{align} 
    z_{k} = \sqrt{\frac{m\omega}{2}}\left( q_{k} + i\,\frac{p_{k}}{m\omega} \right).
\end{align}
Expressed in terms of such canonical coordinates, the Hamilton equations then follow as
\begin{subequations}
\label{eq:qp_dot_recip}
\begin{align}
    &\dot{q}_{k} = \sqrt{\frac{2}{m\omega}} \Re{\dot{z}_{k}} = \frac{p_{k}}{m} + \pdv{V_{\mathrm{q-c}}}{p_{k}},\\
    &\dot{p}_{k} = \sqrt{2m\omega} \Im{\dot{z}_{k}} = -m\omega^{2}q_{k} - \pdv{V_{\mathrm{q-c}}}{q_{k}},
\end{align}
\end{subequations}
which is indeed in agreement with Paper I \cite{alex2021}.

Notably, from the appearance of gradient contributions to \textit{both} $\dot{q}_{k}$ and $\dot{p}_{k}$, it can be clearly seen that while the involved canonical coordinates play the role of ``position'' and ``momentum'' within the Hamilton equations, they are not to be associated with physical positions and momenta \cite{alex2021}.
This is further reflected by the coordinate rescaling applied upon a switch in FSSH. This rescaling is obtained by taking the real and imaginary parts of Eq.~\ref{eq:rescale}, yielding
\begin{subequations}
\begin{align}
    &p_{k}' = p_{k} - \gamma \Braket{\tilde{\alpha} | \frac{\partial}{\partial q_{k}} | \tilde{\beta}}, \\ &q_{k}' = q_{k} + \gamma \Braket{\tilde{\alpha} | \frac{\partial}{\partial p_{k}} | \tilde{\beta}},
\end{align}
\end{subequations}
in agreement with Paper II \cite{alex2022}.

In Papers I \cite{alex2021} and II \cite{alex2022}, application of Eq.~\ref{eq:qp_dot_recip} in conjunction with a reciprocal-space description of the quantum subsystem was shown to yield results in agreement with those obtained fully within a physical basis. This is a direct consequence of the formal equivalence between the Hamilton equations expressed in the physical basis, Eq.~\ref{eq:HamiltonsEOM}, and the reciprocal-space variant, Eq.~\ref{eq:qp_dot_recip}.

For pristine lattices, electronic wavepackets tend to localize in low-energy regions of reciprocal space, allowing a reciprocal-space quantum basis to be truncated to those regions \cite{qiu2016, tempelaar2019}. Moreover, since phonons act so as to absorb or emit lattice momentum during electron--phonon scattering events, a reciprocal-space classical basis can be truncated to select regions in conjunction with quantum basis truncations \cite{alex2021, alex2022}. Such truncations cannot be performed in the physical basis, since electronic carriers and phonons are both delocalized over the entire lattice. Importantly, such truncations offer radical savings of the computational cost of simulating the electron--phonon scattering dynamics.

\section{Application to an impurity model}\label{sec:impurity}

While the reciprocal-space treatment of pristine lattices presented in Papers I \cite{alex2021} and II \cite{alex2022} provides a compelling demonstration of the utility of transformed MQC dynamics, it is the generalized equations of motion presented in the present Article that allows optimal basis transformations to be applied even when the lattice is disrupted. In such cases, lattice momentum is no longer a good quantum number, compromising the effectiveness of reciprocal-space MQC dynamics and prompting the need for alternative representations in order to enable effective basis truncations. To demonstrate this, we proceed to present results for a lattice involving a single impurity.

\subsection{Model}\label{sec:model}

\begin{figure}[t!]
    \centering
    \includegraphics{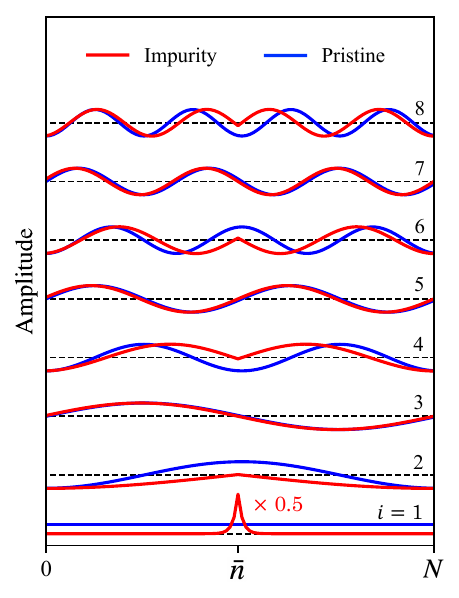}
    \caption{Amplitudes of the 8 lowest purely-electronic eigenfunctions of a pristine lattice (blue) and an impurity model (red). The impurity site is taken to be $\bar{n} = N/2$, and site $n=0$ corresponds to $n=N$ due to periodic boundaries. Amplitudes are shown with arbitrary (but constant) scaling and are offset vertically for visual clarity. An additional $\times\;0.5$ scaling is applied to the $i=1$ amplitude for the impurity model. Corresponding calculations invoked a total of $N=100$ lattice sites.}
    \label{fig:eigenfunctions}
\end{figure}

As in Paper I \cite{alex2021}, we consider a single electronic carrier while representing the lattice by a tight-binding model. The associated purely-electronic quantum Hamiltonian is given by
\begin{align}
    \hat{H}_{\mathrm{q}} = 
    -J \sum_{n} \left( \hat{c}_{n+1}^\dagger \hat{c}_{n} + \hat{c}_{n}^\dagger \hat{c}_{n+1}\right) - \Delta \,\hat{c}_{\bar{n}}^\dagger \hat{c}_{\bar{n}}.
\end{align}
Here, $\hat{c}_{n}^\dagger$ and $\hat{c}_{n}$ represent the annihilation and creation operators for the electronic carrier, respectively, associated with lattice site $n$, and $J$ is the nearest-neighbor interaction term. Furthermore, $\bar{n}$ denotes the impurity site and $\Delta$ is the associated energetic detuning relative to the other lattice sites. Periodic boundaries are imposed, so that $n=N$ corresponds to $n=0$. This Hamiltonian is solved for in order to obtain the purely-electronic eigenstates obeying,
\begin{align}
    \hat{H}_{\mathrm{q}} \ket{\phi_i} = E_i \ket{\phi_i}.
\end{align}
Here and throughout, $i$ is used to label the purely-electronic eigenstates, which are not to be confused with the instantaneous eigenstates of the total quantum Hamiltonian (cf.~Eq.~\ref{eq:SE_full}), which are denoted by $\alpha$ instead. While the instantaneous eigenstates depend parametrically on the classical coordinates, thereby attaining a time-dependence, the purely-electronic eigenstates form a time-independent, diabatic basis. As a convention, $i$ is taken to run with increasing eigenenergy.

Shown in Fig.~\ref{fig:eigenfunctions} are the purely-electronic eigenfunction amplitudes within the physical basis for a pristine lattice with $J = 1.0$ and $\Delta = 0$ and for an impurity model with the detuning was adjusted to $\Delta = 2.0$. Here and throughout, we take parameter values to be unitless. We note, however, that, when taking the thermal energy at room temperature (293 K) as a reference, a unit of energy amounts to 25 meV. As can be seen in Fig.~\ref{fig:eigenfunctions}, for the pristine lattice, the eigenstates assume the periodic oscillatory profiles indicative of Bloch states. For the impurity model, however, the lowest-energy eigenstate (with $i = 1$) is largely localized on the impurity site, $\bar{n}$. All other eigenstates with odd $i$ values ($i = 3, 5, 7, \ldots$) reproduce the Bloch-like states (save for a slight and arbitrary phase shift). For even $i$, the eigenstates tend to approach the Bloch-like states with increasing values of $i$, although maintaining a deviation close to $\bar{n}$. In the course of nonadiabatic dynamics, eigenstate population is expected to funnel from the delocalized Bloch-like states towards the localized state at lowest energy. It is this lowest-energy state that is challenging to resolve within a reciprocal space basis, as a localized state involves contributions from many reciprocal-space basis states due to the underlying Fourier relationship.

In order to find an alternative basis within which to effectively represent the nonadiabatic dynamics of the impurity model, we first observe that the previously-applied reciprocal-space representation \cite{alex2021, alex2022} effectively invokes the purely-electronic eigenstates of the pristine lattice, i.e., the eigenstates of $\hat{H}_\mathrm{q}$ shown in Fig.~\ref{fig:eigenfunctions}. It should be noted that the eigenstates of the pristine lattice shown in this figure were solved for through a real-valued eigenvalue decomposition of the real-valued Hamiltonian $\hat{H}_{\mathrm{q}}$. This yields cosine and sine solutions rather than the complex exponents commonly appearing in the Bloch formalism. However, one can trivially transform between both solutions by taking symmetric and antisymmetric combinations of degenerate eigenstates. Under Bloch-like solutions, truncations of the reciprocal-space basis were previously invoked by introducing a wavevector cutoff, denoted $k_0$, such that basis states having $\vert k\vert>k_0$ were excluded \cite{alex2021, alex2022}. For the real-valued solutions shown in Fig.~\ref{fig:eigenfunctions}, it is more appropriate to introduce an energy cutoff, $E_\mathrm{c}$, such that basis states are excluded having $E_i>E_\mathrm{c}$. Importantly, this effectively yields identical basis truncations, since the Bloch-state energies increase monotonically with $\abs{k}$ (provided that $J>0$).

While the reciprocal-space basis may not offer much benefit to describing the impurity model, the notion of using the purely-electronic eigenbasis, which differs between the pristine lattice and the impurity model, is an interesting choice to consider. Indeed, the purely-electronic eigensolutions of the impurity model tend to simultaneously capture the extendedness and localization necessary to describe the nonadiabatic scattering onto the impurity, as shown in Fig.~\ref{fig:eigenfunctions}. Moreover, the underlying transformation also captures the localized nature of the nuclear vibration responsible for self-trapping at the impurity site, while simultaneously representing the (approximately) momentum-carrying phonons throughout the rest of the lattice. For that reason, in describing the impurity model, we will resort to the purely-electronic eigenstates as a basis for both the quantum \emph{and} classical subsystems.

We should stress that this choice of basis is heuristic, and that superior bases are likely to exist. Moreover, there is no need to keep with the same basis when describing the quantum and classical subsystems. Identifying the theoretically-optimal bases is not a trivial task, however, and we reserve a thorough exploration of this topic for a follow-up study. Instead, by adopting the purely-electronic eigenbasis, we will present a proof-of-principles of the general applicability of transformed MQC dynamics, and the possibilities for basis truncations it affords. Within the purely-electronic eigenbasis representation for the impurity model, we invoke basis truncations similarly to that applied to the real-valued solutions of the pristine lattice, by introducing an energy cutoff, $E_\mathrm{c}$.

As an initial condition of the quantum system, we consider the single-carrier excitation with zero lattice momentum, given by
\begin{align}
    \label{eq:psi_initial}
    \ket{\Psi_0} = \ket{k=0} = \frac{1}{\sqrt{N}} \sum_{n} \ket{n}.
\end{align}
This initial condition is representative of a tightly-bound electron--hole pair (Frenkel exciton) produced upon impulsive optical excitation (under the long-wavelength approximation).

We invoke the Holstein model in order to account for the nuclear modes driving the nonadiabatic dynamics. Accordingly, the classical subsystem is taken to consist of identical harmonic modes, such that $V_\mathrm{c}(\{q_{n}\})$ is given by Eq.~\ref{eq:V_harmonic} and $z_n$ is given by Eq.~\ref{eq:z_n_harmonic}.
The operator governing the quantum--classical interactions is then given by
\begin{align}
    \hat{H}_{\mathrm{q-c}} = g \sqrt{2\omega^3} \sum_{n} \hat{c}_{n}^\dagger \hat{c}_{n} q_n.
\end{align}
In what follows, we adopt the parameters from Fig.~\ref{fig:eigenfunctions}, and additionally set $\omega = 0.2$ and $g = 1.0$, while adjusting the total number of lattice sites to $N=30$. The classical coordinates $q_n$ and $p_n$ are initially and independently drawn from a Boltzmann distribution \cite{alex2021} at a temperature $T = 1.0$.
We reiterate that, when taking the thermal energy at room temperature (293 K) as a reference, a unit of energy amounts to 25 meV. By the same token, a unit of time amounts to 164 fs.

\subsection{Results}\label{sec:results}

In the following, we will present FSSH calculations for both the case of a pristine lattice, by setting $\Delta = 0$, and that of a single impurity model, by setting $\Delta = 2.0$. With the pristine lattice, we will be revisiting a system that was addressed in Papers I \cite{alex2021} and II \cite{alex2022}, but with adjusted parameters. The FSSH calculations presented were performed within the reciprocal-space basis as well as in the purely-electronic eigenbasis. For the former, real-valued eigenvector projections (cf.~Sec.~\ref{sec:MQC}), necessary for momentum rescalings, were obtained following the procedure outlined in Paper II \cite{alex2022}. For the purely-electronic eigenbasis, all relevant eigenvectors are real-valued by construction, and were taken as is. Both treatments are fully consistent with our recently-proposed gauge-invariant generalization \cite{krotz2024}.

The electronic populations were evaluated by first constructing the total electronic density matrix within the instantaneous eigenbasis of the total Hamiltonian operator, $\hat{H}_\mathrm{q} + \hat{H}_\mathrm{q-c}$. As is commonly done \cite{tempelaar2013surface, landry2013communication, chen2016on}, the diagonal elements of this density matrix were based on the active surfaces, while the offdiagonal elements were based on the electronic wavefunction coefficients, i.e.,
\begin{align}
    \rho_{\alpha\beta} = \delta_{\alpha\beta} \delta_{\alpha a} + (1 - \delta_{\alpha\beta}) A_{\alpha}^* A_{\beta},
\end{align}
where $\delta$ is the Kronecker delta function. This density matrix was then transformed \cite{tempelaar2017, alex2022} to the physical basis, reciprocal-space basis, and purely-electronic eigenbasis, respectively, after which corresponding populations were obtained through $P_n = \rho_{nn}$, $P_k = \rho_{kk}$, and $P_i = \rho_{ii}$.

\begin{figure}[t!]
    \centering
    \includegraphics{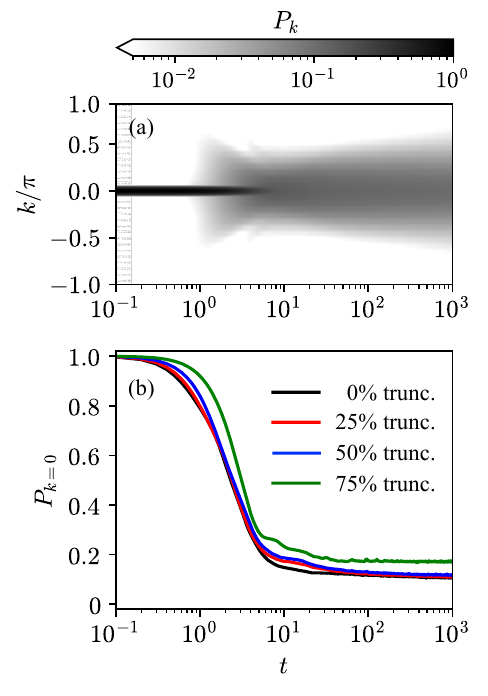}
    \caption{(a) Transient electronic populations $P_k$ calculated within reciprocal-space FSSH for a pristine ($\Delta = 0$) lattice with $N=30$ sites, and with $J=1.0$, $\omega=0.2$, $g=1.0$, and $T=1.0$. (b) Dynamics of $P_{k=0}$ under varying truncations of the reciprocal-space basis. Percentages of truncations shown were reached by varying $k_0$ (see text).}
    \label{fig:t1.0_pristine_dynamics}
\end{figure}

Shown in Fig.~\ref{fig:t1.0_pristine_dynamics} are results for the pristine lattice. Here, only reciprocal-space FSSH was applied, recognizing the formal equivalence with the purely-electronic eigenbasis in this limit (as discussed in Sec.~\ref{sec:MQC}). Fig.~\ref{fig:t1.0_pristine_dynamics} (a) presents time-dependent reciprocal-space electronic populations, $P_k$, obtained without any basis truncation imposed. As seen here, and as discussed in Papers I \cite{alex2021} and II \cite{alex2022}, a scattering of the electronic carrier out of the $k=0$ initial state is observed.

Fig.~\ref{fig:t1.0_pristine_dynamics} (b) depicts the time-dependent zero-momentum electronic population, $P_{k=0}$, resulting from untruncated calculations, together with $P_{k=0}$ calculated under increasing basis truncations. This population is seen to remain invariant under truncations of up to 50\% of the reciprocal-space basis, which is consistent with the findings of Paper II \cite{alex2022}, where FSSH was applied to a pristine lattice under Holstein-type coupling between the carrier and the nuclear modes. This is rationalized by the carrier being both initiated and thermally-biased towards $k=0$, with high-momentum basis states providing only small contributions to the nonadiabatic dynamics.

\begin{figure*}[t!]
    \centering    
    \includegraphics[width=6.5in]{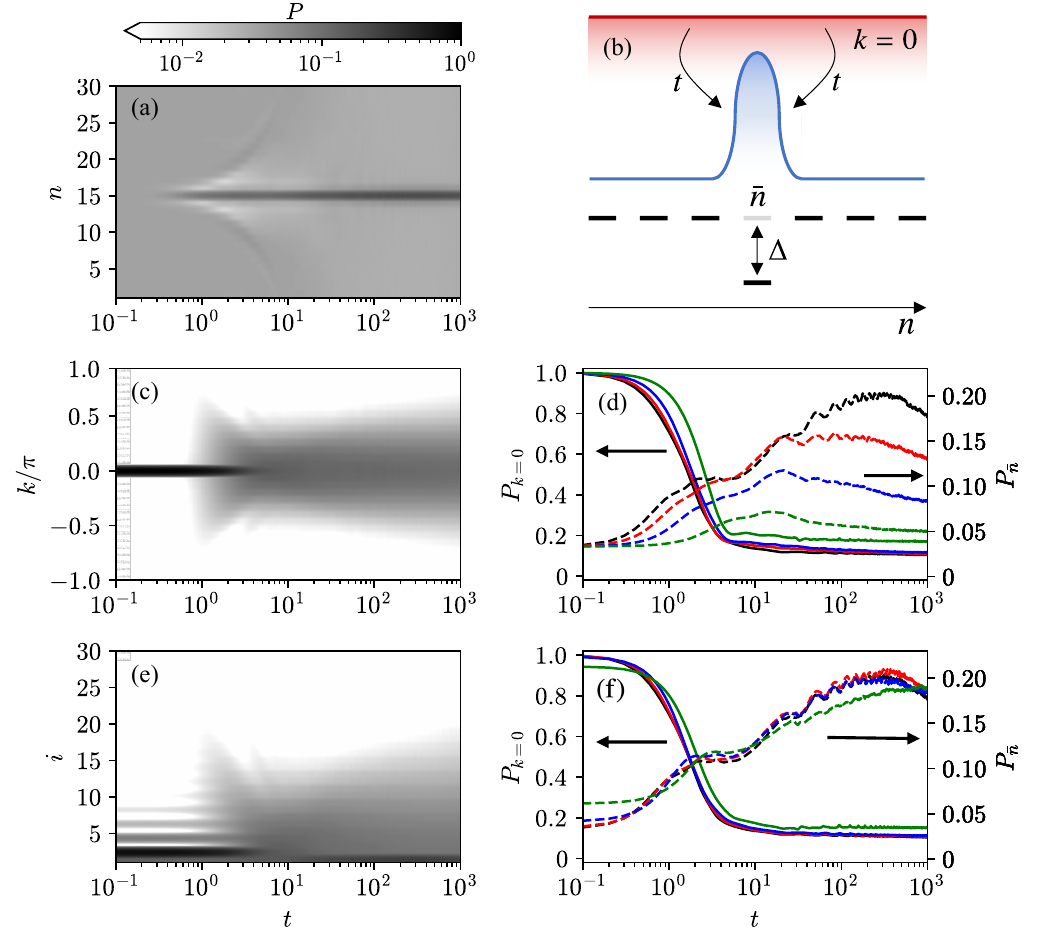}
    \caption{(a) Transient electronic populations $P_n$ calculated by conventional FSSH formulated within the physical basis for an impurity model. Parameters are as in Fig.~\ref{fig:t1.0_pristine_dynamics}, except for $\Delta = 2.0$. (b) Schematic of the population dynamics shown in (a), where an initial $k=0$ excitation funnels towards a state localized at the impurity site, $n=\bar{n}$. (c) $P_k$ calculated by reciprocal-space FSSH. (d) Corresponding populations $P_{k=0}$ (solid, left axis) and $P_{\bar{n}}$ (dashed, right axis) under varying truncations of the reciprocal-space basis. Curve colorings are as in Fig.~\ref{fig:t1.0_pristine_dynamics} (b), while percentages of truncations were reached by varying $k_0$. (e) $P_i$ calculated by FSSH formulated within the purely-electronic eigenbasis. (f) Same as (e), but under varying truncations of the purely-electronic eigenbasis. Curve colorings are as in Fig.~\ref{fig:t1.0_pristine_dynamics} (b), while percentages of truncations were reached by varying $E_\mathrm{c}$ (see text).}
    \label{fig:t1.0_imp2_collection}
\end{figure*}

A similar analysis is presented in Fig.~\ref{fig:t1.0_imp2_collection}, but for the impurity model subjected to FSSH within reciprocal space and within the purely-electronic eigenbasis. Shown as a reference in Fig.~\ref{fig:t1.0_imp2_collection} (a) are time-dependent populations $P_n$, obtained through FSSH in the physical basis. Here, the $k = 0$ initial state is seen to be fully delocalized, as expected from Eq.~\ref{eq:psi_initial}, while nonadiabatic dynamics funnels the carrier into the impurity with $n = \bar{n}$. A schematic illustration of this process is depicted in Fig.~\ref{fig:t1.0_imp2_collection} (b). Shown in Fig.~\ref{fig:t1.0_imp2_collection} (c) and (e) are $P_k$ and $P_i$, respectively, as obtained through FSSH within the reciprocal-space basis and within the purely-electronic eigenbasis. While the dynamics of $P_k$ appears to be quite similar to that shown for the pristine lattice in Fig.~\ref{fig:t1.0_pristine_dynamics} (a), $P_i$ clearly exhibits the gradual trapping of excitation in the lowest-energy eigenstate with $i = 1$.

We will proceed to assess to what extent truncations of the reciprocal-space basis and the purely-electronic eigenbasis allows capturing of both the delocalization of the initial excitation as well as the transient localization at the impurity site. To this end, we evaluate $P_{k=0}$ as well as $P_{\bar{n}}$. Fig.~\ref{fig:t1.0_imp2_collection} (d) and (f) show both populations as obtained through FSSH under truncations of the reciprocal-space basis and of the purely-electronic basis, respectively. As can be seen here, both truncated bases perform well in describing $P_{k=0}$. A markedly-worse performance is found for $P_{\bar{n}}$ under reciprocal-space basis truncations, which is a direct consequence of reciprocal space providing a poor representation of local phenomena. Due to the Fourier relationship, accounting for the local population at site $\bar{n}$ requires the complete reciprocal-space basis to be included, and truncations of this basis yield a proportional drop in this site population.
    
The purely-electronic eigenbasis, however, is found to perform remarkably well in describing $P_{\bar{n}}$ under truncations of up to 50\%. This is the result of the lowest-energy eigenstate capturing most of the local population at site $\bar{n}$. As such, the truncated purely-electronic eigenbasis is shown to outperform the truncated reciprocal-space basis in governing the funneling of electronic population into the impurity, as we anticipated based on Fig.~\ref{fig:eigenfunctions}.

\section{Conclusions and outlook}\label{sec:conclusions}

In summary, we have derived a formulation of MQC dynamics within arbitrary bases for the classical and quantum coordinates. This allows any given system to be optimally treated by finding preferred bases which can be efficiently truncated while retaining good agreement with untruncated calculations. Such is demonstrated by our application of FSSH to a system involving an electronic carrier scattering onto a single impurity in an otherwise pristine lattice while interacting with phonons. The funneling from an initial, delocalized state towards a low-energy state localized at the impurity was faithfully captured when transforming the quantum and classical coordinates into the purely-electronic eigenbasis, followed by significant basis truncations. A similar level of truncation within the reciprocal-space basis instead yielded significant distortions of the impurity population.

The encouraging results obtained for the impurity model offer promising prospects for the realistic application of MQC dynamics, and in particular FSSH, to materials involving disruptions of the crystal lattice. In recent years, FSSH has found applications to materials involving an impurity \cite{zhang2018}, defects \cite{chu2020b}, and wrinkling \cite{xu2023}, as well as materials where the lattice symmetry was disrupted by nearby molecules \cite{chu2016}. While no basis transformations were invoked in these studies, our present analysis suggest them to offer significant gains in performance and scalability when combined with efficient basis truncations.
    
We re-emphasize that the purely-electronic eigenbasis adopted for the impurity model was chosen based entirely on heuristic arguments and, in spite of its good performance, may not be the theoretically-optimal choice. Finding the optimal choice for a given system is nontrivial, and as such we reserve such inquiry for future research. Here, it should be noted that the bases adopted for the quantum and classical subsystems need not be the same, as was adopted in the present Article. Rather both bases can be independently optimized, resorting to the full special unitary group for each \cite{krotz2024}.

Importantly, the transformed equations of motion presented in this Article are applicable to the full suite of MQC methods, beyond FSSH. An example of particular interest is Ehrenfest dynamics, which is unrivaled in terms of simplicity and computational affordability. These favorable attributes notwithstanding, Ehrenfest dynamics is known to suffer from over-thermalization in the asymptotic time limit \cite{parandekar2005mixed, parandekar2006detailed}. In addition to contributing to the method's inaccuracy, this deteriorates the effectiveness of basis truncations, as quantum excitations tend to delocalize over an excess number of basis states \cite{alex2022}. We should emphasize that the intrinsic accuracy of MQC modeling is not addressed in the present Article. In Paper II \cite{alex2022}, we found the results generated by FSSH for a lattice model to show good agreement with full-quantum reference calculations, yet much remains to be learned about the performance of FSSH in materials modeling. In that regard, a recently-proposed coherent generalization of FSSH is noteworthy \cite{tempelaar2017, bondarenko2023}, which may offer particular advantages in capturing the coherent dynamics prevalent in materials.
    
Lastly, we note that the transformed classical dynamics derived in Sec.~\ref{sec:classical} is completely general. Its ability to incorporate mode anharmonicities, although not taken advantage of in our current demonstration of the impurity model, enables broad application to molecular dynamics simulations, with or without accompanying quantum modeling. Subjecting a given classical system to unitary basis transformations, the same way quantum-mechanical systems are treated, may provide a straightforward route to optimally representing its dynamics by a truncated set of coordinates.

\begin{acknowledgement}
This material is based upon work supported by the National Science Foundation under Grant No.~2145433. K.M.~gratefully acknowledges support from the Mark A.~Ratner Postdoctoral Fellowship and the Northwestern University International Institute for Nanotechnology (IIN).
\end{acknowledgement}

\bibliography{bibliography}

\providecommand{\latin}[1]{#1}
\makeatletter
\providecommand{\doi}
  {\begingroup\let\do\@makeother\dospecials
  \catcode`\{=1 \catcode`\}=2 \doi@aux}
\providecommand{\doi@aux}[1]{\endgroup\texttt{#1}}
\makeatother
\providecommand*\mcitethebibliography{\thebibliography}
\csname @ifundefined\endcsname{endmcitethebibliography}  {\let\endmcitethebibliography\endthebibliography}{}
\begin{mcitethebibliography}{46}
\providecommand*\natexlab[1]{#1}
\providecommand*\mciteSetBstSublistMode[1]{}
\providecommand*\mciteSetBstMaxWidthForm[2]{}
\providecommand*\mciteBstWouldAddEndPuncttrue
  {\def\EndOfBibitem{\unskip.}}
\providecommand*\mciteBstWouldAddEndPunctfalse
  {\let\EndOfBibitem\relax}
\providecommand*\mciteSetBstMidEndSepPunct[3]{}
\providecommand*\mciteSetBstSublistLabelBeginEnd[3]{}
\providecommand*\EndOfBibitem{}
\mciteSetBstSublistMode{f}
\mciteSetBstMaxWidthForm{subitem}{(\alph{mcitesubitemcount})}
\mciteSetBstSublistLabelBeginEnd
  {\mcitemaxwidthsubitemform\space}
  {\relax}
  {\relax}

\bibitem[Taghizadeh \latin{et~al.}(2017)Taghizadeh, Hipolito, and Pedersen]{taghizadeh2017}
Taghizadeh,~A.; Hipolito,~F.; Pedersen,~T.~G. Linear and nonlinear optical response of crystals using length and velocity gauges: Effect of basis truncation. \emph{Phys. Rev. B} \textbf{2017}, \emph{96}, 195413\relax
\mciteBstWouldAddEndPuncttrue
\mciteSetBstMidEndSepPunct{\mcitedefaultmidpunct}
{\mcitedefaultendpunct}{\mcitedefaultseppunct}\relax
\EndOfBibitem
\bibitem[Taylor \latin{et~al.}(2020)Taylor, Mandal, Zhou, and Huo]{taylor2020}
Taylor,~M. A.~D.; Mandal,~A.; Zhou,~W.; Huo,~P. Resolution of Gauge Ambiguities in Molecular Cavity Quantum Electrodynamics. \emph{Phys. Rev. Lett.} \textbf{2020}, \emph{125}, 123602\relax
\mciteBstWouldAddEndPuncttrue
\mciteSetBstMidEndSepPunct{\mcitedefaultmidpunct}
{\mcitedefaultendpunct}{\mcitedefaultseppunct}\relax
\EndOfBibitem
\bibitem[Bloch(1929)]{bloch1929}
Bloch,~F. {\"U}ber die Quantenmechanik der Elektronen in Kristallgittern. \emph{Zeitschrift f{\"u}r Physik} \textbf{1929}, \emph{52}, 555--600\relax
\mciteBstWouldAddEndPuncttrue
\mciteSetBstMidEndSepPunct{\mcitedefaultmidpunct}
{\mcitedefaultendpunct}{\mcitedefaultseppunct}\relax
\EndOfBibitem
\bibitem[Kittel(1966)]{kittel1966}
Kittel,~C. \emph{Introduction to Solid State Physics}; John Wiley \& Sons, 1966\relax
\mciteBstWouldAddEndPuncttrue
\mciteSetBstMidEndSepPunct{\mcitedefaultmidpunct}
{\mcitedefaultendpunct}{\mcitedefaultseppunct}\relax
\EndOfBibitem
\bibitem[REDFIELD(1965)]{redfield1965}
REDFIELD,~A. In \emph{Advances in Magnetic Resonance}; Waugh,~J.~S., Ed.; Advances in Magnetic and Optical Resonance; Academic Press, 1965; Vol.~1; pp 1--32\relax
\mciteBstWouldAddEndPuncttrue
\mciteSetBstMidEndSepPunct{\mcitedefaultmidpunct}
{\mcitedefaultendpunct}{\mcitedefaultseppunct}\relax
\EndOfBibitem
\bibitem[Tempelaar and Reichman(2018)Tempelaar, and Reichman]{tempelaar2018}
Tempelaar,~R.; Reichman,~D.~R. {Vibronic exciton theory of singlet fission. III. How vibronic coupling and thermodynamics promote rapid triplet generation in pentacene crystals}. \emph{The Journal of Chemical Physics} \textbf{2018}, \emph{148}, 244701\relax
\mciteBstWouldAddEndPuncttrue
\mciteSetBstMidEndSepPunct{\mcitedefaultmidpunct}
{\mcitedefaultendpunct}{\mcitedefaultseppunct}\relax
\EndOfBibitem
\bibitem[Nelson \latin{et~al.}(2014)Nelson, Fernandez-Alberti, Roitberg, and Tretiak]{nelson2014}
Nelson,~T.; Fernandez-Alberti,~S.; Roitberg,~A.~E.; Tretiak,~S. Nonadiabatic Excited-State Molecular Dynamics: Modeling Photophysics in Organic Conjugated Materials. \emph{Accounts of Chemical Research} \textbf{2014}, \emph{47}, 1155--1164\relax
\mciteBstWouldAddEndPuncttrue
\mciteSetBstMidEndSepPunct{\mcitedefaultmidpunct}
{\mcitedefaultendpunct}{\mcitedefaultseppunct}\relax
\EndOfBibitem
\bibitem[Subotnik \latin{et~al.}(2016)Subotnik, Jain, Landry, Petit, Ouyang, and Bellonzi]{subotnik2016}
Subotnik,~J.~E.; Jain,~A.; Landry,~B.; Petit,~A.; Ouyang,~W.; Bellonzi,~N. Understanding the Surface Hopping View of Electronic Transitions and Decoherence. \emph{Annual Review of Physical Chemistry} \textbf{2016}, \emph{67}, 387--417\relax
\mciteBstWouldAddEndPuncttrue
\mciteSetBstMidEndSepPunct{\mcitedefaultmidpunct}
{\mcitedefaultendpunct}{\mcitedefaultseppunct}\relax
\EndOfBibitem
\bibitem[Wang \latin{et~al.}(2016)Wang, Akimov, and Prezhdo]{wang2016}
Wang,~L.; Akimov,~A.; Prezhdo,~O.~V. Recent Progress in Surface Hopping: 2011--2015. \emph{The Journal of Physical Chemistry Letters} \textbf{2016}, \emph{7}, 2100--2112\relax
\mciteBstWouldAddEndPuncttrue
\mciteSetBstMidEndSepPunct{\mcitedefaultmidpunct}
{\mcitedefaultendpunct}{\mcitedefaultseppunct}\relax
\EndOfBibitem
\bibitem[Crespo-Otero and Barbatti(2018)Crespo-Otero, and Barbatti]{crespo-otero2018}
Crespo-Otero,~R.; Barbatti,~M. Recent Advances and Perspectives on Nonadiabatic Mixed Quantum--Classical Dynamics. \emph{Chemical Reviews} \textbf{2018}, \emph{118}, 7026--7068\relax
\mciteBstWouldAddEndPuncttrue
\mciteSetBstMidEndSepPunct{\mcitedefaultmidpunct}
{\mcitedefaultendpunct}{\mcitedefaultseppunct}\relax
\EndOfBibitem
\bibitem[Nelson \latin{et~al.}(2020)Nelson, White, Bjorgaard, Sifain, Zhang, Nebgen, Fernandez-Alberti, Mozyrsky, Roitberg, and Tretiak]{nelson2020}
Nelson,~T.~R.; White,~A.~J.; Bjorgaard,~J.~A.; Sifain,~A.~E.; Zhang,~Y.; Nebgen,~B.; Fernandez-Alberti,~S.; Mozyrsky,~D.; Roitberg,~A.~E.; Tretiak,~S. Non-adiabatic Excited-State Molecular Dynamics: Theory and Applications for Modeling Photophysics in Extended Molecular Materials. \emph{Chemical Reviews} \textbf{2020}, \emph{120}, 2215--2287\relax
\mciteBstWouldAddEndPuncttrue
\mciteSetBstMidEndSepPunct{\mcitedefaultmidpunct}
{\mcitedefaultendpunct}{\mcitedefaultseppunct}\relax
\EndOfBibitem
\bibitem[Krotz \latin{et~al.}(2021)Krotz, Provazza, and Tempelaar]{alex2021}
Krotz,~A.; Provazza,~J.; Tempelaar,~R. A reciprocal-space formulation of mixed quantum–classical dynamics. \emph{J. Chem. Phys.} \textbf{2021}, \emph{154}, 224101\relax
\mciteBstWouldAddEndPuncttrue
\mciteSetBstMidEndSepPunct{\mcitedefaultmidpunct}
{\mcitedefaultendpunct}{\mcitedefaultseppunct}\relax
\EndOfBibitem
\bibitem[Nie \latin{et~al.}(2014)Nie, Long, Sun, Huang, Zhang, Xiong, Hewak, Shen, Prezhdo, and Loh]{nie2014}
Nie,~Z.; Long,~R.; Sun,~L.; Huang,~C.-C.; Zhang,~J.; Xiong,~Q.; Hewak,~D.~W.; Shen,~Z.; Prezhdo,~O.~V.; Loh,~Z.-H. Ultrafast Carrier Thermalization and Cooling Dynamics in Few-Layer MoS2. \emph{ACS Nano} \textbf{2014}, \emph{8}, 10931--10940\relax
\mciteBstWouldAddEndPuncttrue
\mciteSetBstMidEndSepPunct{\mcitedefaultmidpunct}
{\mcitedefaultendpunct}{\mcitedefaultseppunct}\relax
\EndOfBibitem
\bibitem[Nie \latin{et~al.}(2015)Nie, Long, Teguh, Huang, Hewak, Yeow, Shen, Prezhdo, and Loh]{nie2015}
Nie,~Z.; Long,~R.; Teguh,~J.~S.; Huang,~C.-C.; Hewak,~D.~W.; Yeow,~E. K.~L.; Shen,~Z.; Prezhdo,~O.~V.; Loh,~Z.-H. Ultrafast Electron and Hole Relaxation Pathways in Few-Layer MoS2. \emph{The Journal of Physical Chemistry C} \textbf{2015}, \emph{119}, 20698--20708\relax
\mciteBstWouldAddEndPuncttrue
\mciteSetBstMidEndSepPunct{\mcitedefaultmidpunct}
{\mcitedefaultendpunct}{\mcitedefaultseppunct}\relax
\EndOfBibitem
\bibitem[Long and Prezhdo(2016)Long, and Prezhdo]{long2016}
Long,~R.; Prezhdo,~O.~V. Quantum Coherence Facilitates Efficient Charge Separation at a MoS2/MoSe2 van der Waals Junction. \emph{Nano Letters} \textbf{2016}, \emph{16}, 1996--2003\relax
\mciteBstWouldAddEndPuncttrue
\mciteSetBstMidEndSepPunct{\mcitedefaultmidpunct}
{\mcitedefaultendpunct}{\mcitedefaultseppunct}\relax
\EndOfBibitem
\bibitem[Chu \latin{et~al.}(2016)Chu, Saidi, Zheng, Xie, Lan, Prezhdo, Petek, and Zhao]{chu2016}
Chu,~W.; Saidi,~W.~A.; Zheng,~Q.; Xie,~Y.; Lan,~Z.; Prezhdo,~O.~V.; Petek,~H.; Zhao,~J. Ultrafast Dynamics of Photongenerated Holes at a CH3OH/TiO2 Rutile Interface. \emph{Journal of the American Chemical Society} \textbf{2016}, \emph{138}, 13740--13749\relax
\mciteBstWouldAddEndPuncttrue
\mciteSetBstMidEndSepPunct{\mcitedefaultmidpunct}
{\mcitedefaultendpunct}{\mcitedefaultseppunct}\relax
\EndOfBibitem
\bibitem[Zheng \latin{et~al.}(2018)Zheng, Xie, Lan, Prezhdo, Saidi, and Zhao]{zheng2018}
Zheng,~Q.; Xie,~Y.; Lan,~Z.; Prezhdo,~O.~V.; Saidi,~W.~A.; Zhao,~J. Phonon-coupled ultrafast interlayer charge oscillation at van der Waals heterostructure interfaces. \emph{Phys. Rev. B} \textbf{2018}, \emph{97}, 205417\relax
\mciteBstWouldAddEndPuncttrue
\mciteSetBstMidEndSepPunct{\mcitedefaultmidpunct}
{\mcitedefaultendpunct}{\mcitedefaultseppunct}\relax
\EndOfBibitem
\bibitem[Shi \latin{et~al.}(2020)Shi, Prezhdo, Zhao, and Saidi]{shi2020}
Shi,~Y.; Prezhdo,~O.~V.; Zhao,~J.; Saidi,~W.~A. Iodine and Sulfur Vacancy Cooperation Promotes Ultrafast Charge Extraction at MAPbI3/MoS2 Interface. \emph{ACS Energy Letters} \textbf{2020}, \emph{5}, 1346--1354\relax
\mciteBstWouldAddEndPuncttrue
\mciteSetBstMidEndSepPunct{\mcitedefaultmidpunct}
{\mcitedefaultendpunct}{\mcitedefaultseppunct}\relax
\EndOfBibitem
\bibitem[Smith and Akimov(2019)Smith, and Akimov]{smith2019}
Smith,~B.; Akimov,~A.~V. Modeling nonadiabatic dynamics in condensed matter materials: some recent advances and applications. \emph{Journal of Physics: Condensed Matter} \textbf{2019}, \emph{32}, 073001\relax
\mciteBstWouldAddEndPuncttrue
\mciteSetBstMidEndSepPunct{\mcitedefaultmidpunct}
{\mcitedefaultendpunct}{\mcitedefaultseppunct}\relax
\EndOfBibitem
\bibitem[Zhang \latin{et~al.}(2021)Zhang, Hou, Jiang, Yang, Saidi, Prezhdo, and Li]{zhang2021}
Zhang,~P.; Hou,~Z.; Jiang,~L.; Yang,~J.; Saidi,~W.~A.; Prezhdo,~O.~V.; Li,~W. Weak Anharmonicity Rationalizes the Temperature-Driven Acceleration of Nonradiative Dynamics in Cu2ZnSnS4 Photoabsorbers. \emph{ACS Applied Materials \& Interfaces} \textbf{2021}, \emph{13}, 61365--61373\relax
\mciteBstWouldAddEndPuncttrue
\mciteSetBstMidEndSepPunct{\mcitedefaultmidpunct}
{\mcitedefaultendpunct}{\mcitedefaultseppunct}\relax
\EndOfBibitem
\bibitem[Xie \latin{et~al.}(2022)Xie, Xu, Wang, and Zhuang]{xie2022}
Xie,~H.; Xu,~X.; Wang,~L.; Zhuang,~W. {Surface hopping dynamics in periodic solid-state materials with a linear vibronic coupling model}. \emph{The Journal of Chemical Physics} \textbf{2022}, \emph{156}, 154116\relax
\mciteBstWouldAddEndPuncttrue
\mciteSetBstMidEndSepPunct{\mcitedefaultmidpunct}
{\mcitedefaultendpunct}{\mcitedefaultseppunct}\relax
\EndOfBibitem
\bibitem[Lively \latin{et~al.}(2024)Lively, Sato, Albareda, Rubio, and Kelly]{lively2024}
Lively,~K.; Sato,~S.~A.; Albareda,~G.; Rubio,~A.; Kelly,~A. Revealing ultrafast phonon mediated inter-valley scattering through transient absorption and high harmonic spectroscopies. \emph{Phys. Rev. Res.} \textbf{2024}, \emph{6}, 013069\relax
\mciteBstWouldAddEndPuncttrue
\mciteSetBstMidEndSepPunct{\mcitedefaultmidpunct}
{\mcitedefaultendpunct}{\mcitedefaultseppunct}\relax
\EndOfBibitem
\bibitem[Mahan(2013)]{mahan2013}
Mahan,~G.~D. \emph{Many-particle physics}; Springer Science \& Business Media, 2013\relax
\mciteBstWouldAddEndPuncttrue
\mciteSetBstMidEndSepPunct{\mcitedefaultmidpunct}
{\mcitedefaultendpunct}{\mcitedefaultseppunct}\relax
\EndOfBibitem
\bibitem[Krotz and Tempelaar(2022)Krotz, and Tempelaar]{alex2022}
Krotz,~A.; Tempelaar,~R. A reciprocal-space formulation of surface hopping. \emph{J. Chem. Phys.} \textbf{2022}, \emph{156}, 024105\relax
\mciteBstWouldAddEndPuncttrue
\mciteSetBstMidEndSepPunct{\mcitedefaultmidpunct}
{\mcitedefaultendpunct}{\mcitedefaultseppunct}\relax
\EndOfBibitem
\bibitem[Tully(1990)]{tully1990}
Tully,~J.~C. Molecular dynamics with electronic transitions. \emph{J. Chem. Phys.} \textbf{1990}, \emph{93}, 1061--1071\relax
\mciteBstWouldAddEndPuncttrue
\mciteSetBstMidEndSepPunct{\mcitedefaultmidpunct}
{\mcitedefaultendpunct}{\mcitedefaultseppunct}\relax
\EndOfBibitem
\bibitem[Chen \latin{et~al.}(2024)Chen, Wang, and Dou]{chen2024}
Chen,~J.; Wang,~Y.; Dou,~W. Floquet Nonadiabatic Mixed Quantum-Classical Dynamics in Laser-Dressed Solid Systems. \emph{arXiv preprint arXiv:2402.12732} \textbf{2024}, \relax
\mciteBstWouldAddEndPunctfalse
\mciteSetBstMidEndSepPunct{\mcitedefaultmidpunct}
{}{\mcitedefaultseppunct}\relax
\EndOfBibitem
\bibitem[Kim \latin{et~al.}(2022)Kim, Aydin, Daza, Avanaki, Keski-Rahkonen, and Heller]{kim2022}
Kim,~D.; Aydin,~A.; Daza,~A.; Avanaki,~K.~N.; Keski-Rahkonen,~J.; Heller,~E.~J. Coherent charge carrier dynamics in the presence of thermal lattice vibrations. \emph{Phys. Rev. B} \textbf{2022}, \emph{106}, 054311\relax
\mciteBstWouldAddEndPuncttrue
\mciteSetBstMidEndSepPunct{\mcitedefaultmidpunct}
{\mcitedefaultendpunct}{\mcitedefaultseppunct}\relax
\EndOfBibitem
\bibitem[Hammes‐Schiffer and Tully(1994)Hammes‐Schiffer, and Tully]{sharon1994}
Hammes‐Schiffer,~S.; Tully,~J.~C. Proton transfer in solution: Molecular dynamics with quantum transitions. \emph{J. Chem. Phys.} \textbf{1994}, \emph{101}, 4657--4667\relax
\mciteBstWouldAddEndPuncttrue
\mciteSetBstMidEndSepPunct{\mcitedefaultmidpunct}
{\mcitedefaultendpunct}{\mcitedefaultseppunct}\relax
\EndOfBibitem
\bibitem[Miao \latin{et~al.}(2019)Miao, Bellonzi, and Subotnik]{miao2019}
Miao,~G.; Bellonzi,~N.; Subotnik,~J. An Extension of the Fewest Switches Surface Hopping Algorithm to Complex {{Hamiltonians}} and Photophysics in Magnetic Fields: {{Berry}} Curvature and ``Magnetic'' Forces. \emph{J. Chem. Phys.} \textbf{2019}, \emph{150}, 124101\relax
\mciteBstWouldAddEndPuncttrue
\mciteSetBstMidEndSepPunct{\mcitedefaultmidpunct}
{\mcitedefaultendpunct}{\mcitedefaultseppunct}\relax
\EndOfBibitem
\bibitem[Wu and Subotnik(2021)Wu, and Subotnik]{wu2021}
Wu,~Y.; Subotnik,~J.~E. Semiclassical Description of Nuclear Dynamics Moving through Complex-Valued Single Avoided Crossings of Two Electronic States. \emph{J. Chem. Phys.} \textbf{2021}, \emph{154}, 234101\relax
\mciteBstWouldAddEndPuncttrue
\mciteSetBstMidEndSepPunct{\mcitedefaultmidpunct}
{\mcitedefaultendpunct}{\mcitedefaultseppunct}\relax
\EndOfBibitem
\bibitem[Bian \latin{et~al.}(2022)Bian, Wu, Teh, and Subotnik]{bian2022}
Bian,~X.; Wu,~Y.; Teh,~H.-H.; Subotnik,~J.~E. Incorporating {{Berry Force Effects}} into {{The Fewest Switches Surface Hopping Algorithm}}: {{Intersystem Crossing}} and {{The Case}} of {{Electronic Degeneracy}}. \emph{J. Chem. Theory Comput.} \textbf{2022}, \emph{18}, 2075--2090\relax
\mciteBstWouldAddEndPuncttrue
\mciteSetBstMidEndSepPunct{\mcitedefaultmidpunct}
{\mcitedefaultendpunct}{\mcitedefaultseppunct}\relax
\EndOfBibitem
\bibitem[Daggett \latin{et~al.}(2024)Daggett, Yang, Liu, and Muechler]{daggett2024}
Daggett,~C.; Yang,~K.; Liu,~C.-X.; Muechler,~L. Toward a Topological Classification of Nonadiabaticity in Chemical Reactions. \emph{Chem. Mater.} \textbf{2024}, \emph{36}, 3479--3489\relax
\mciteBstWouldAddEndPuncttrue
\mciteSetBstMidEndSepPunct{\mcitedefaultmidpunct}
{\mcitedefaultendpunct}{\mcitedefaultseppunct}\relax
\EndOfBibitem
\bibitem[Krotz and Tempelaar(2024)Krotz, and Tempelaar]{krotz2024}
Krotz,~A.; Tempelaar,~R. Treating geometric phase effects in nonadiabatic dynamics. \emph{Phys. Rev. A} \textbf{2024}, \emph{109}, 032210\relax
\mciteBstWouldAddEndPuncttrue
\mciteSetBstMidEndSepPunct{\mcitedefaultmidpunct}
{\mcitedefaultendpunct}{\mcitedefaultseppunct}\relax
\EndOfBibitem
\bibitem[Qiu \latin{et~al.}(2016)Qiu, da~Jornada, and Louie]{qiu2016}
Qiu,~D.~Y.; da~Jornada,~F.~H.; Louie,~S.~G. Screening and many-body effects in two-dimensional crystals: Monolayer ${\mathrm{MoS}}_{2}$. \emph{Phys. Rev. B} \textbf{2016}, \emph{93}, 235435\relax
\mciteBstWouldAddEndPuncttrue
\mciteSetBstMidEndSepPunct{\mcitedefaultmidpunct}
{\mcitedefaultendpunct}{\mcitedefaultseppunct}\relax
\EndOfBibitem
\bibitem[Tempelaar and Berkelbach(2019)Tempelaar, and Berkelbach]{tempelaar2019}
Tempelaar,~R.; Berkelbach,~T.~C. Many-body simulation of two-dimensional electronic spectroscopy of excitons and trions in monolayer transition metal dichalcogenides. \emph{Nature Communications} \textbf{2019}, \emph{10}, 3419\relax
\mciteBstWouldAddEndPuncttrue
\mciteSetBstMidEndSepPunct{\mcitedefaultmidpunct}
{\mcitedefaultendpunct}{\mcitedefaultseppunct}\relax
\EndOfBibitem
\bibitem[Tempelaar \latin{et~al.}(2013)Tempelaar, van~der Vegte, Knoester, and Jansen]{tempelaar2013surface}
Tempelaar,~R.; van~der Vegte,~C.~P.; Knoester,~J.; Jansen,~T. L.~C. {Surface hopping modeling of two-dimensional spectra}. \emph{The Journal of Chemical Physics} \textbf{2013}, \emph{138}, 164106\relax
\mciteBstWouldAddEndPuncttrue
\mciteSetBstMidEndSepPunct{\mcitedefaultmidpunct}
{\mcitedefaultendpunct}{\mcitedefaultseppunct}\relax
\EndOfBibitem
\bibitem[Landry \latin{et~al.}(2013)Landry, Falk, and Subotnik]{landry2013communication}
Landry,~B.~R.; Falk,~M.~J.; Subotnik,~J.~E. {Communication: The correct interpretation of surface hopping trajectories: How to calculate electronic properties}. \emph{The Journal of Chemical Physics} \textbf{2013}, \emph{139}, 211101\relax
\mciteBstWouldAddEndPuncttrue
\mciteSetBstMidEndSepPunct{\mcitedefaultmidpunct}
{\mcitedefaultendpunct}{\mcitedefaultseppunct}\relax
\EndOfBibitem
\bibitem[Chen and Reichman(2016)Chen, and Reichman]{chen2016on}
Chen,~H.-T.; Reichman,~D.~R. {On the accuracy of surface hopping dynamics in condensed phase non-adiabatic problems}. \emph{The Journal of Chemical Physics} \textbf{2016}, \emph{144}, 094104\relax
\mciteBstWouldAddEndPuncttrue
\mciteSetBstMidEndSepPunct{\mcitedefaultmidpunct}
{\mcitedefaultendpunct}{\mcitedefaultseppunct}\relax
\EndOfBibitem
\bibitem[Tempelaar and Reichman(2017)Tempelaar, and Reichman]{tempelaar2017}
Tempelaar,~R.; Reichman,~D.~R. {Generalization of fewest-switches surface hopping for coherences}. \emph{The Journal of Chemical Physics} \textbf{2017}, \emph{148}, 102309\relax
\mciteBstWouldAddEndPuncttrue
\mciteSetBstMidEndSepPunct{\mcitedefaultmidpunct}
{\mcitedefaultendpunct}{\mcitedefaultseppunct}\relax
\EndOfBibitem
\bibitem[Zhang \latin{et~al.}(2018)Zhang, Zheng, Xie, Lan, Prezhdo, Saidi, and Zhao]{zhang2018}
Zhang,~L.; Zheng,~Q.; Xie,~Y.; Lan,~Z.; Prezhdo,~O.~V.; Saidi,~W.~A.; Zhao,~J. Delocalized Impurity Phonon Induced Electron--Hole Recombination in Doped Semiconductors. \emph{Nano Letters} \textbf{2018}, \emph{18}, 1592--1599\relax
\mciteBstWouldAddEndPuncttrue
\mciteSetBstMidEndSepPunct{\mcitedefaultmidpunct}
{\mcitedefaultendpunct}{\mcitedefaultseppunct}\relax
\EndOfBibitem
\bibitem[Chu \latin{et~al.}(2020)Chu, Saidi, Zhao, and Prezhdo]{chu2020b}
Chu,~W.; Saidi,~W.~A.; Zhao,~J.; Prezhdo,~O.~V. Soft Lattice and Defect Covalency Rationalize Tolerance of $\beta$-CsPbI3 Perovskite Solar Cells to Native Defects. \emph{Angewandte Chemie International Edition} \textbf{2020}, \emph{59}, 6435--6441\relax
\mciteBstWouldAddEndPuncttrue
\mciteSetBstMidEndSepPunct{\mcitedefaultmidpunct}
{\mcitedefaultendpunct}{\mcitedefaultseppunct}\relax
\EndOfBibitem
\bibitem[Xu \latin{et~al.}(2023)Xu, Zhou, Alexeev, Cadore, Paradisanos, Ott, Soavi, Tongay, Cerullo, Ferrari, Prezhdo, and Loh]{xu2023}
Xu,~C.; Zhou,~G.; Alexeev,~E.~M.; Cadore,~A.~R.; Paradisanos,~I.; Ott,~A.~K.; Soavi,~G.; Tongay,~S.; Cerullo,~G.; Ferrari,~A.~C.; Prezhdo,~O.~V.; Loh,~Z.-H. Ultrafast Electronic Relaxation Dynamics of Atomically Thin MoS2 Is Accelerated by Wrinkling. \emph{ACS Nano} \textbf{2023}, \emph{17}, 16682--16694\relax
\mciteBstWouldAddEndPuncttrue
\mciteSetBstMidEndSepPunct{\mcitedefaultmidpunct}
{\mcitedefaultendpunct}{\mcitedefaultseppunct}\relax
\EndOfBibitem
\bibitem[Parandekar and Tully(2005)Parandekar, and Tully]{parandekar2005mixed}
Parandekar,~P.~V.; Tully,~J.~C. Mixed quantum-classical equilibrium. \emph{J. Chem. Phys.} \textbf{2005}, \emph{122}, 094102\relax
\mciteBstWouldAddEndPuncttrue
\mciteSetBstMidEndSepPunct{\mcitedefaultmidpunct}
{\mcitedefaultendpunct}{\mcitedefaultseppunct}\relax
\EndOfBibitem
\bibitem[Parandekar and Tully(2006)Parandekar, and Tully]{parandekar2006detailed}
Parandekar,~P.~V.; Tully,~J.~C. Detailed Balance in Ehrenfest Mixed Quantum-Classical Dynamics. \emph{J. Chem. Theory Comput.} \textbf{2006}, \emph{2}, 229--235\relax
\mciteBstWouldAddEndPuncttrue
\mciteSetBstMidEndSepPunct{\mcitedefaultmidpunct}
{\mcitedefaultendpunct}{\mcitedefaultseppunct}\relax
\EndOfBibitem
\bibitem[Bondarenko and Tempelaar(2023)Bondarenko, and Tempelaar]{bondarenko2023}
Bondarenko,~A.~S.; Tempelaar,~R. {Overcoming positivity violations for density matrices in surface hopping}. \emph{The Journal of Chemical Physics} \textbf{2023}, \emph{158}, 054117\relax
\mciteBstWouldAddEndPuncttrue
\mciteSetBstMidEndSepPunct{\mcitedefaultmidpunct}
{\mcitedefaultendpunct}{\mcitedefaultseppunct}\relax
\EndOfBibitem
\end{mcitethebibliography}

\end{document}